# On droplet falling velocity

Wenjie Ji[1], Siyuan Wang[2], Jiguang Hao[1*], and J. M. Floryan[3]

[1]School of Aerospace Engineering, Beijing Institute of Technology, Beijing 100081, China
[2]China Academy of Launch Vehicle Technology, Beijing 100076, China
[3]Department of Mechanical and Materials Engineering, Western University, London, Ontario N6A 5B9, Canada

Droplet velocities used in impact studies were investigated using high-speed photography. It was determined that droplets do not reach terminal velocity before a typical impact, raising the question of how to predict impact velocity. This question was investigated experimentally, and the results were used to validate a theoretical model. Experiments used droplets with diameters 0.70mm to 4.0mm, liquids with a density of 791kg/m$^3$ to 1261.5kg/m$^3$, and viscosities 1.0mPa s to 1390.0mPa s, release height up to 1.0m. The ambient pressure was varied between atmospheric and 25kPa. It was shown that the droplet velocity increased with the droplet diameter, liquid density, release height, and ambient pressure reduction but changed marginally with viscosity. A simple dynamic model accounting for the aerodynamic drag was proposed. This model, which uses empirical formulae to determine the instantaneous drag coefficient, predicts velocity, which agrees well with the experimental data within the range of parameters used in this study. It provides a valuable tool for the design of droplet impact studies.

## 1. Introduction

Droplets are widely encountered both in nature and in a variety of applications. Examples include rain[1-3], printing[4-6], combustion[7-10], cooling[11, 12], coating[13, 14], aerosol formation[15], virus spreading[16-19], pesticide delivery[11], extraction[20] and additive manufacturing[21-23]. Generally, a droplet impacts its target at a specific speed which determines the impact outcomes, i.e., splashing [24-26], spreading [27, 28], or penetration [29-31]. No general theoretical method exists to determine the impact velocity as the droplet accelerates just before the impact. As a result, its determination relies on experiments required in various applications, e.g., experimental studies on droplet impact [25, 30, 32-34], virus spreading[17, 18], printing[5, 6], which limits the ability to predict the outcome of droplet impact, e.g., difficulties in determination of safe social distancing. The droplet terminal velocities are well known, but variations of the velocity during the acceleration process are yet to be investigated in detail. As acceleration is rarely finalized in impacts found in applications, this contribution aims to provide a useful theoretical tool for determining the droplet velocity during different stages of its acceleration.

The droplet's acceleration process results from an interplay of its weight, aerodynamic drag, and inertia. The aerodynamic resistance is given as $F_d = 1/8\, \rho_g V^2 C_d \pi D^2$ [35] where $C_d$ is the drag coefficient, $\rho_g$ is the air density, $V$ is the droplet velocity, and $D$ is the droplet diameter. For small droplet velocity, $F_d$ is dominated by friction [36]. As the velocity increases, the aerodynamic (pressure) resistance increases in a complex manner, leading to difficulties in theoretical prediction of the droplet velocity, especially when the pressure is varied[37-40].

It is simple to establish a model to determine the droplet velocity[41, 42] using Newton's second law if $C_d$ can be determined accurately[43]. For a spherical body, approximation $C_d = 0.5$ is generally acceptable for small enough $Re$'s [44] with deviations increasing as $Re$ increases[45]. For very low Reynolds number $Re = \rho_g V D/\mu_g < 0.1$, $C_d$ can be determined using Stokes' law[46, 47]. Here, $\mu_g$ is the air viscosity. $C_d$ strongly depended on $Re$ for higher $Re$ due to the asymmetry of the pressure field between the upstream and downstream sides of the droplet, with this dependence usually determined experimentally[48]. Various empirical and semi-empirical formulas have been proposed for a wide range of $Re$'s[45, 48-61] However, it is still unclear which formula best predicts the velocity of an accelerating droplet. Additional complications arise for droplets larger than the capillary length as the shape deformation influences $C_d$ [62].

Here, we experimentally determined droplet velocity just before its impact as a function of its release height,

diameter, liquid density and viscosity, and ambient air pressure. A simple model to predict this velocity was proposed utilizing an instantaneous $C_d$. The model was validated through comparison with experiments. It is shown that using an adequately determined $C_d$ leads to a highly accurate prediction of the droplet velocity, providing a useful theoretical tool for its determination under various conditions of interest in droplet impact applications.

## 2. Experiment

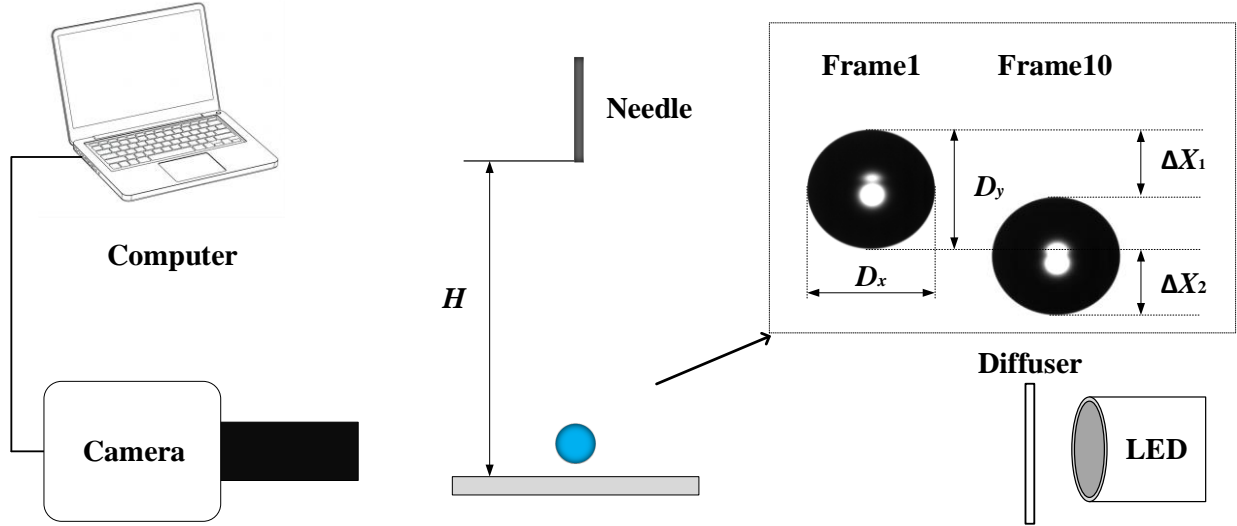

**Fig. 1.** Sketch of the experimental apparatus. $H$ is the release height. The process used to determine the droplet velocity and diameter is illustrated in the inset. $D_x$ and $D_y$ are the horizontal and vertical droplet diameters, respectively. $\Delta X_1$ and $\Delta X_2$ are the displacements of the upper and lower droplet edges occurring between the tenth and first image frames before the impact, respectively.

As shown in Fig. 1, droplets with diameters ranging from 0.7 to 4.0 mm were generated either using a flat-tipped needle driven by a syringe pump (for $D > 1.8$ mm, dripping method) or using a homemade DOD device [30] (for $D < 1.8$ mm). A piezoelectric ceramic buzzer drives the DOD and can produce droplets with diameters ranging from 0.7 to 1.5 mm; see [30] for detailed information. Experiments used deionized water, ethanol, and glycerin, with their properties summarized in Table I.

Table I. Physical properties of experimental liquids ($T = 23\pm1$°C)

| Liquid | $\rho$ (kg/m$^3$) | $\sigma$ (mNm$^{-1}$) | $\mu$ (mPa s) |
|---|---|---|---|
| Water | 998.7 | 72.9 | 1.00 |
| Ethanol | 791.0 | 22.9 | 1.19 |
| Glycerin | 1261.5 | 63.3 | 1390.00 |

The release height $H$ was adjusted in a range of 0.1 to 1.0 m for experiments carried out at ambient pressure while in a range of 0.1 to 0.78 m for experiments at reduced pressures. A transparent vacuum chamber controlled the pressure of 10-101 kPa[25, 39].

A Photron Nova S12 high-speed camera was used to record the droplet movement and diameter at a rate of 20,000 fps and with a spatial resolution of 12.4 μm /pixel. LEDs combined with a diffuser were used to provide illumination. Experiments were repeated three times for each set of conditions listed in Table II.

Table II. Experimental conditions
set 1: $H$ (m) = 0.1;0.2;0.3;0.4;0.5;0.6;0.8;1.0; set 2: $H$ (m) = 0.1;0.2;0.3;0.4;0.5;0.6;0.78

| Number | $D$ (mm) | Liquid | $P$ (kPa) | set |
|---|---|---|---|---|
| 1 | 0.70 | Water | 101 | 1 |
| 2 | 0.93 | Water | 101 | 1 |
| 3 | 1.08 | Water | 101 | 1 |
| 4 | 1.50 | Water | 101 | 1 |
| 5 | 2.30 | Water | 101 | 1 |
| 6 | 2.74 | Water | 101 | 1 |
| 7 | 3.40 | Water | 101 | 1 |
| 8 | 4.00 | Water | 101 | 1 |
| 9 | 2.30 | Ethanol | 101 | 1 |
| 10 | 2.30 | Glycerin | 101 | 1 |
| 11 | 1.84 | Ethanol | 25/50/75 | 2 |
| 12 | 2.30 | Water | 25/50/75 | 2 |
| 13 | 2.74 | Water | 25/50/75 | 2 |
| 14 | 4.00 | Water | 25/50/75 | 2 |

The droplet diameter was calculated[36, 55, 56] as

$$D = (D_x^2 D_y)^{(1/3)} \quad (1)$$

where $D_x$ and $D_y$ are the horizontal and vertical droplet diameters, respectively; see the inset in Fig. 1. All reported diameter values represent averages from multiple experiments. The droplet velocity was calculated as

$$V = \frac{\Delta X}{\Delta T} \quad (2)$$

where $\Delta X = (\Delta X_1 + \Delta X_2)/2$, $\Delta X_1$ and $\Delta X_2$ are displacements of the upper and lower edges of the droplet between the first and tenth frames 10 the impact (the droplet impacts the plate at frame 11), respectively (see the inset in Fig.1) and $\Delta T = 0.0005$s is the time for the droplet to move between these frames. The velocities reported here represent averages of three experiments under identical conditions.

## 3. Theoretical or empirical methods to determine droplet velocity

The acceleration of the droplet is determined by an interplay between its weight, inertia, and aerodynamic resistance. The weight of a spherical droplet is $G = 4/3\pi(D/2)^3 \rho g$, where $g$ stands for the gravitational acceleration and $\rho$ is the density of the liquid. The aerodynamic resistance is $F_d = 1/2\pi \rho_g C_d (D/2)^2 V^2$, where $\rho_g$ is the air density. The droplet acceleration is

$$\frac{dV}{dT} = \frac{(G - F_d)}{4/3\pi(D/2)^3 \rho} = g - \frac{(3C_d \rho_g V^2)}{4D\rho} \quad (3)$$

where $T$ is time measured from droplet release time, and the droplet velocity can be determined using

$$V = \frac{dH}{dT} \quad (4)$$

In Eq. (3), $C_d$ is the only undetermined parameter. Ignoring resistance results in $C_d = 0$ and $V = \sqrt{2gH}$ which provide limiting values for very low ambient pressure.

In general, $C_d$ is not negligible and is given by an empirical formula [45, 48, 55]

$$C_d = \frac{24}{Re}(1 + k(Re)) \quad (5)$$

where $k(R_e)$ is a $Re$–dependent correction factor determined experimentally, typically using wind tunnel experiments with spherical solid models; such experiments do not account for droplet deformation, oscillations, and internal dynamics.

For $Re < 1$, the Stokes law is satisfied[48], i.e., $k(Re) \approx 0$. As $D$ or/and $V$ increases, the pressure effects contribute to the resistance, and the relation $k(Re)$ is usually determined empirically. Fuchs[63] proposed an empirical expression

$$C_d = f_1(Re) = \frac{24}{Re}(1+0.158Re^{2/3}). \tag{6}$$

This relation has been used in determining the droplet velocity[44] in the present study, with variations of $C_d$ illustrated using a solid magenta line in Fig. 2.

Yang[64] collected historical data for $C_d$ for a smooth sphere and used a multigene genetic programming approach to produce correlation (7). $C_d$ prediction based on this correlation yielded the green dashed line in Fig. 2.

$$\begin{cases} C_d = f_2(Re) = R(Re) + err(Re), \\ R(Re) = \frac{24}{Re}(1 + \frac{3/16 Re}{1+(19/240Re)/(1+1/122Re)}), \\ err(Re) = \frac{8}{1000} \frac{ln[(1+Re)/21000][ln((1+Re)/254)]^2}{1+Re/7000}. \end{cases} \tag{7}$$

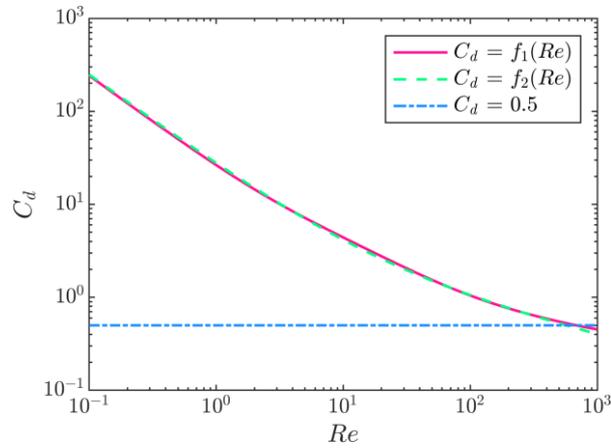

**Fig. 2.** Variations of the drag coefficient $C_d$ determined using Eq. (6) (the solid magenta line) and (7) (the green dashed line) as a function of $Re$. The blue dash-dotted line provides the reference value $C_d = 0.5$.

A SCILAB code was developed to solve numerically Eqs (3) and (4) for $C_d$ either varying with velocity as described by Eq. (6) or (7) or taken as $C_d = 0.5$. This code evaluated instantaneous $C_d$ and produced $V$ as a function of $H$ with $V = 0.0$, $H = 0.0$, and $T = 0.0$ as initial conditions, 0.001s as the time step, and a specific value of $H$ as the end condition.

Previous studies on droplet impacts used an alternative empirical expression (8)[65-69] to determine the droplet velocity for a specific value of $H$ based on its terminal velocity

$$V = V_T [1 - \exp(\frac{-2gH}{V_T^2})]^{\frac{1}{2}} \tag{8}$$

where $V_T$ is a function of droplet diameter and can be determined using the empirical expression[66, 70]:

$$V_T = 9.65 - 10.3\exp(-600D) \tag{9}$$

Predictions based on the five models, i.e., $C_d = f_1(R_e), f_2(R_e), 0, 0.5$, and a combination of Eqs (8) and (9) were compared with the experimental results to determine the most accurate model.

## 4. Results

### 4.1 At ambient pressure

Figure 3 illustrates variations of velocity $V$ of water droplets with diameters of (a) $D$ = 0.70mm, (b) $D$ = 0.93mm, (c) $D$ = 1.08mm, (d) $D$ = 1.5mm, (e) $D$ = 2.3mm, (f) $D$ = 2.74mm, (g) $D$ = 3.4mm, (h) $D$ = 4.0mm as a function of the release height $H$ at ambient pressure. The velocity increases as an increase of $H$ for all droplet diameters. The bigger the droplet, the larger velocity for the same $H$. This trend is increasingly evident with the increase of $H$, indicating that acceleration decreases faster for smaller droplets with an increase of $H$ or $V$.

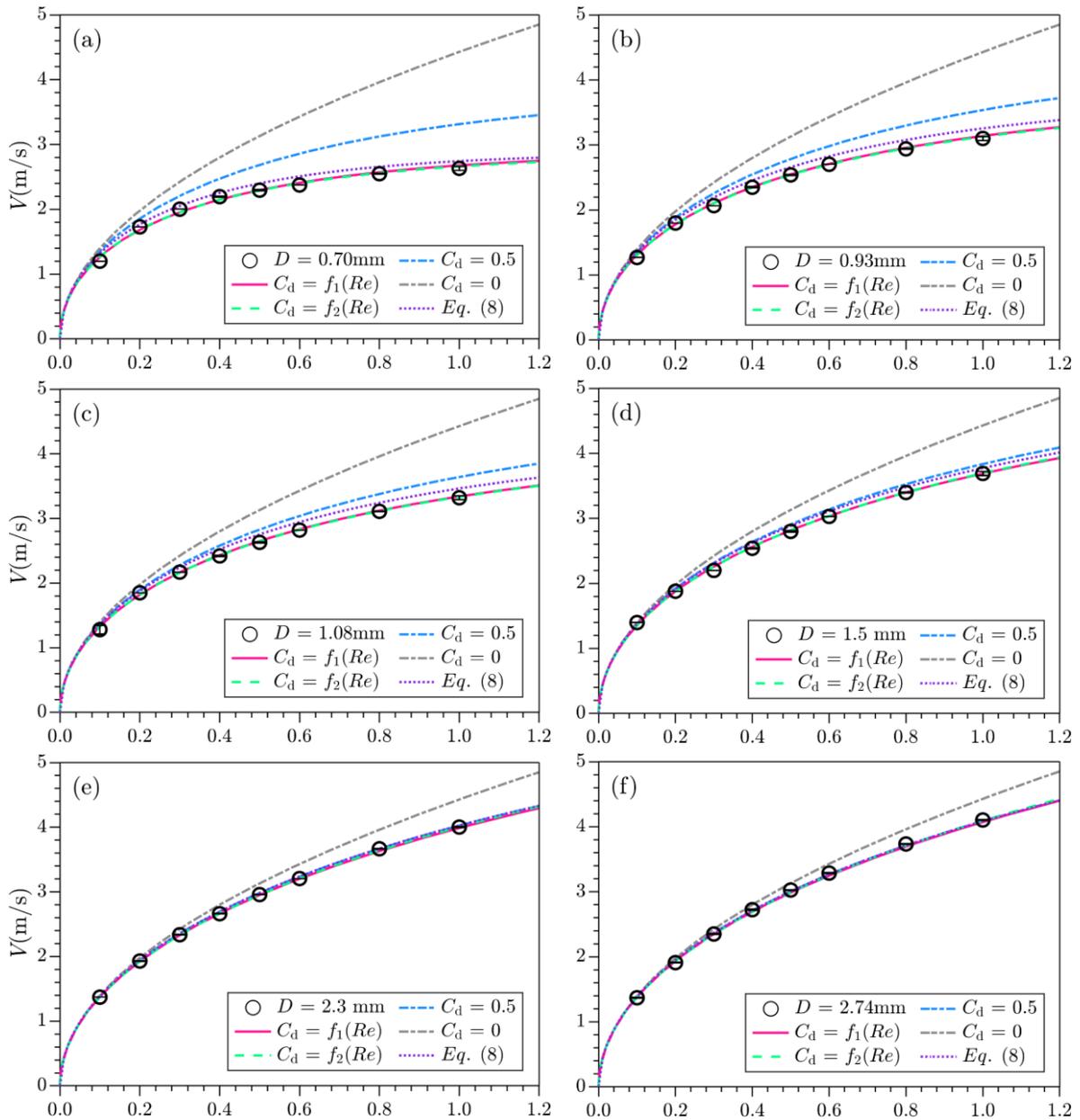

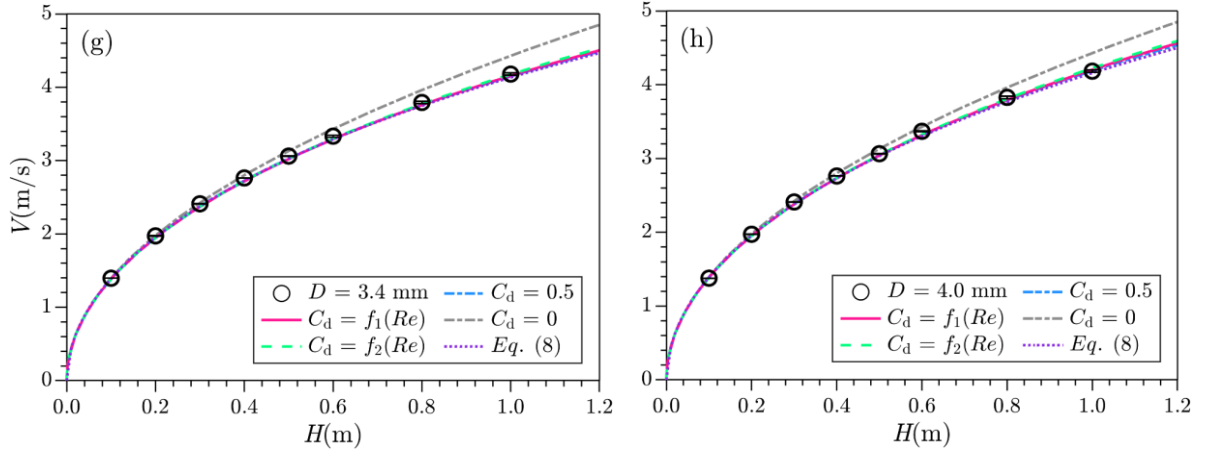

**Fig. 3.** Variations of $V$ as a function of $H$ for water droplets with diameters (a) $D = 0.70$mm, (b) $D = 0.93$mm, (c) $D = 1.08$mm, (d) $D = 1.50$mm, (e) $D = 2.30$mm, (f) $D = 2.74$mm, (g) $D = 3.40$mm, (h) $D = 4.00$mm. Circles identify the experimental points. Error bars indicate the standard deviations. Magenta solid, green dashed, blue, and grey dash-dotted lines represent theoretically determined velocity with $C_d$ determined using Eqs (6) and (7), $C_d = 0.5$, and $C_d = 0$, respectively. The purple dotted line represents velocity determined using Eqs (8) and (9).

Lines in Fig.3 show the theoretically-determined falling velocities as a function of $H$ using the five models discussed above. The magenta solid and green dashed lines illustrate velocities determined by numerically solving Eqs (3) and (4) using instantaneous $C_d$ determined from Eqs (6) and (7), respectively. They both agree well with the experimental results regardless of the diameters. The grey dash-dotted line illustrates velocity predicted using $C_d = 0$, i.e., with negligible air resistance. Comparison with experimental results shows that the air resistance-induced reduction of the falling velocity is more significant for the smaller droplets and the higher the release height. The blue dash-dotted line illustrates velocity determined using $C_d = 0.5$. The predictions for larger droplets, i.e., $D = 2.3$mm, 2.74mm, 3.4mm, and 4.0mm, are similar to those obtained using Eqs (6) and (7). The deviation between predictions based on $C_d = 0.5$ and the experimental results increases with a reduction of the droplet diameter, indicating that $C_d = 0.5$ produces good results only for a limited range of droplet diameter, i.e., 2.3mm $< D <$ 4.0mm. The purple dotted line illustrates velocities predicted using Eqs (8) and (9) – these predictions fall between those obtained using $C_d = 0$ and 0.5 but are less accurate than those determined using instantaneous $C_d$'s described by Eqs. (6) and (7), especially for the four smaller droplets. Results produced by all five methods agree well with the experiments for the lowest release height of $H = 0.1$m (i.e., smallest droplet velocities), indicating that the air resistance can be neglected for small $H$ with $V = \sqrt{2gH}$ providing a reasonably accurate velocity prediction[36]. For larger droplets of $D = 3.4$mm, and 4.0mm, the velocities determined by $V = \sqrt{2gH}$ also agree reasonably with the experiments even for $H = 0.6$ m, indicating that use $V = \sqrt{2gH}$ for velocity predictions is reasonable for large droplet released from $H < 0.6$m[44].

Figure 4 illustrates variations of velocity $V$ for water droplets as a function of the droplet diameter $D$ for the release height $H$ of 0.8 m (a) and 1.0 m (b), respectively. It provides intuitive support to the statement above that $V$ increases as an increase of $D$ for the same $H$, especially for smaller droplets of $D < 2.3$ mm. Predictions based on $C_d$ determined by Eqs (6) and (7) provide the best matching with experimental measurements for all droplet diameters used in this study; predictions based on Eqs (8) and (9) agree with the experiments reasonably. The model using $C_d = 0.5$ produces good results for a droplet diameter range of 2.3mm $< D <$ 4.0 mm with error increasing with a decrease of $D$. For higher $H$'s of 0.8m and 1.0m, the method using $C_d = 0$ cannot produce reasonable predictions, especially for small droplets.

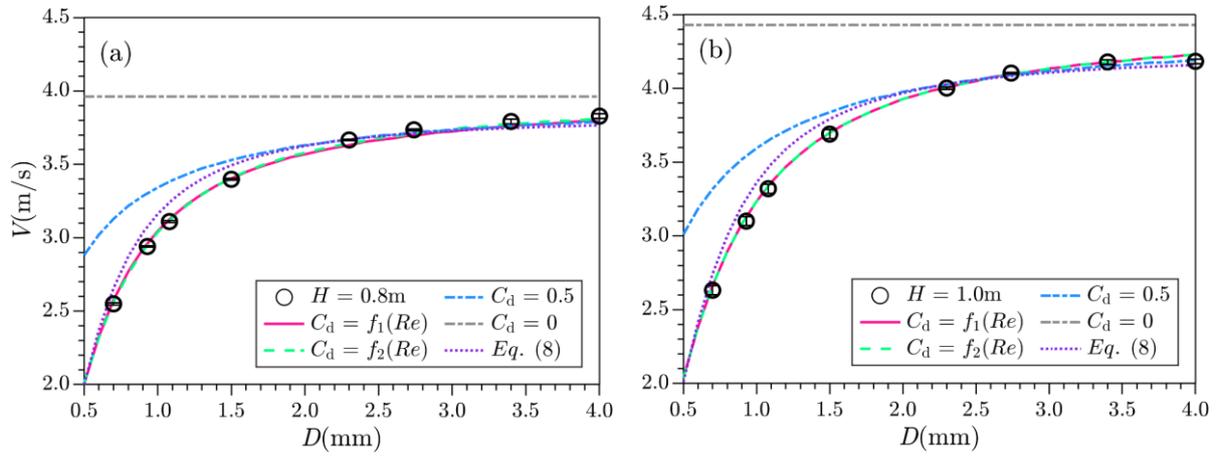

**Fig. 4.** Variations of velocity $V$ as a function of $D$ for $H$ (a) 0.8 m and (b) 1.0 m. Symbols and lines are the same as in Fig. 3.

Figure 5 illustrates variations of $V$ for ethanol (a) and glycerin (b) droplets of $D = 2.30$ mm as a function of the release height $H$. Variations of $V$ determined experimentally for ethanol and glycerin exhibit the same trend as the water droplets. Theoretical predictions based on the numerical solution of Eqs (3) and (4) with different instantaneous $C_d$'s, i.e., 0.5, determined either by Eq. (6) or by Eq. (7), account for droplet density and, thus, produce accurate predictions for droplets made of different liquids. Use of Eqs (8) and (9) produces the same velocities regardless liquid's density; they are larger than those observed experimentally for ethanol but smaller for glycerin droplets.

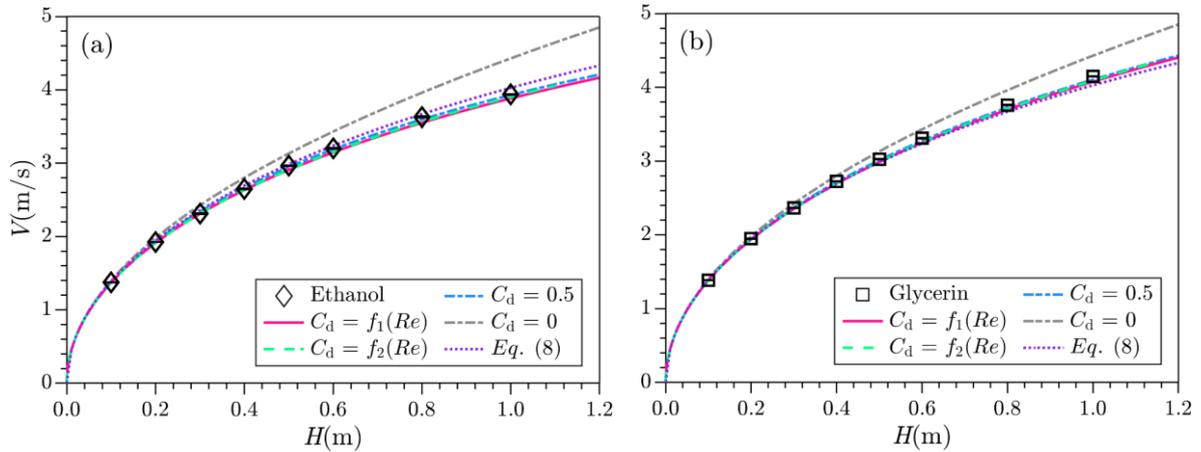

**Fig. 5.** Variations of velocity $V$ as a function of the release height $H$ for (a) ethanol and (b) glycerin droplets of $D = 2.30$ mm. Diamonds identify experimental points for ethanol, and squares identify experimental points for glycerin. Error bars indicate the standard deviations. The lines are the same as in Fig. 3.

Figure 6 illustrates variations of $V$ for (a) $H = 0.8$ m and (b) 1.0 m as a function of liquid density $\rho$ for droplets with $D = 2.30$ mm. Diamonds (ethanol), circles (water), and squares (glycerin) identify experimental points. Lines illustrate theoretical predictions. Results demonstrate that $V$ increases as an increase of $\rho$ for the same $H$ due to a larger weight increase than the air resistance $F_d$. Theoretical predictions based on the first three models agree well with this trend, but predictions based on Eqs (8) and (9) are insensitive to density variations, further indicating the limitations of these models.

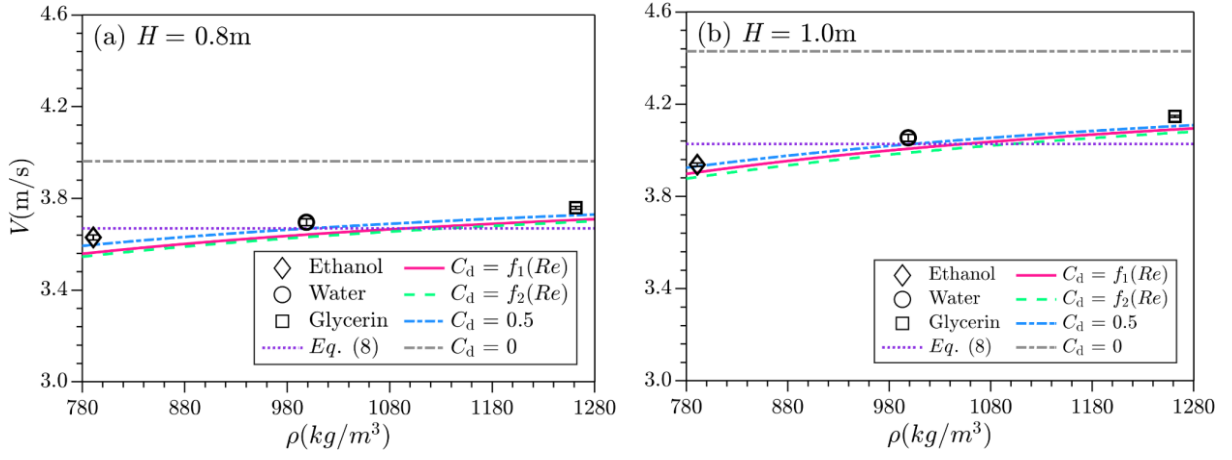

**Fig. 6**. Variations of velocity $V$ as a function of the droplet density $\rho$ for $H$ (a) 0.8m and (b) 1.0m for $D$ = 2.30 mm. Ethanol, water, and glycerin results are identified using diamonds, circles, and squares. Error bars indicate the standard deviations. The lines are the same as in Fig. 3.

**4.2 At reduced pressure**

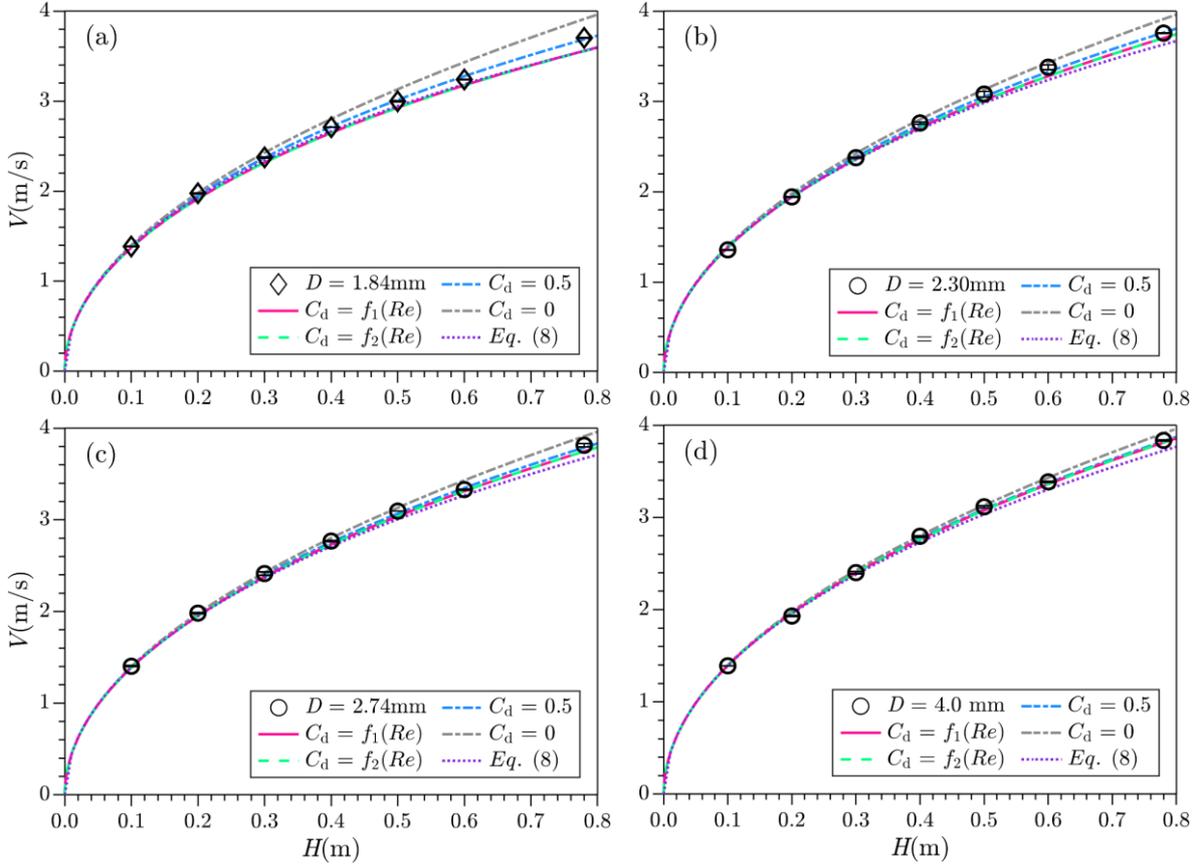

**Fig. 7.** Variations of velocity $V$ as a function of $H$ for droplets of (a) ethanol with $D$ = 1.84mm, water with (b) $D$ = 2.3 mm, (c) $D$ = 2.74mm, (d) $D$ = 4.0 mm at $P$ = 50 kPa. Diamonds and circles identify the experimental points of ethanol and water. Error bars indicate the standard deviations. Lines are as stated in the caption of Fig. 3.

Figure 7 illustrates variations of velocity $V$ of droplets of (a) ethanol with $D$ = 1.84mm, water with (b) $D$ = 2.30mm, (c) $D$ = 2.74mm, and (d) $D$ = 4.00mm as functions $H$ at a reduced pressure of $P$ = 50 kPa. It was not possible to experiment with smaller droplets since the DOD cannot work at reduced pressure due to the formation of air bubbles from dissolved air. Compared with the results at ambient pressure, the results are closer to predictions based on $C_d = 0$ (the grey dash-dotted line), indicating that the reduced pressure decreases the air resistance. To consider the air pressure, the first three models determine the air density using the ideal-gas state equation and produce predictions that agree well with the experiments. Predictions based on Eqs (8) and (9) remain unchanged as these relations cannot account for air pressure variations.

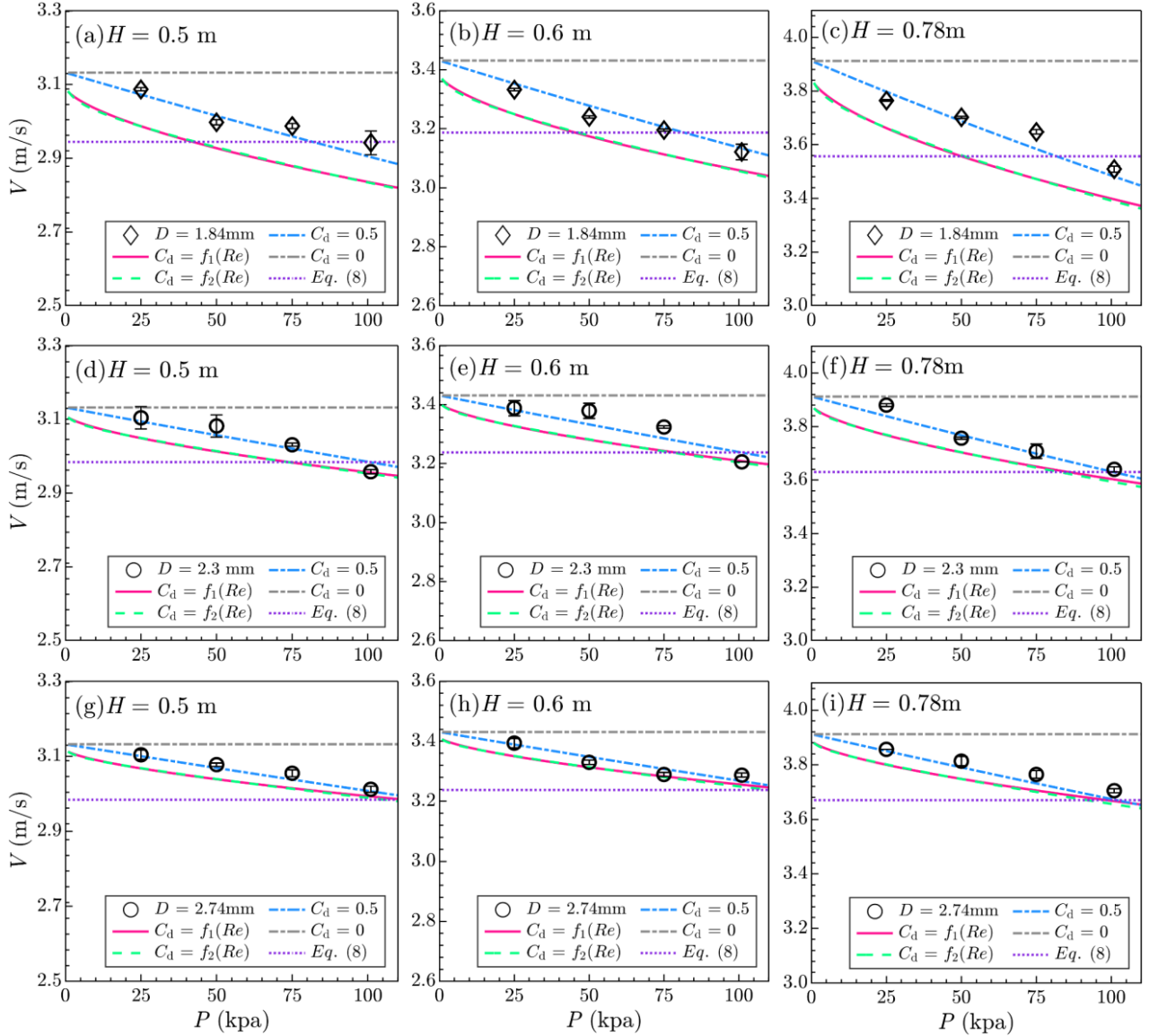

**Fig. 8**. Variations of velocity $V$ as a function of the air pressure $P$ for ethanol droplets with $D = 1.84$mm for $H = 0.5$m (a), 0.6m (b), and 0.78m (c), for water droplets with $D = 2.3$mm for $H = 0.5$m (d), 0.6m (e) and 0.78m (f), and for water droplets with $D = 2.74$mm for $H = 0.5$m (g), 0.6m (h) and 0.78m (i). Diamonds and circles identify the experimental points for ethanol and water. Error bars indicate the standard deviations. Lines are as stated in the caption of Fig. 3.

Figure 8 illustrates variations of velocity $V$ for ethanol droplets with $D = 1.84$ mm released at (a) $H = 0.5$m, (b) $H = 0.6$m, and (c) $H = 0.78$m (c) for water droplets with $D = 2.3$ mm released at (d) $H = 0.5$m, (e) $H = 0.6$m, and (f) $H = 0.78$m, for water droplets with $D = 2.74$ mm released at (g) $H = 0.5$m, (h) $H = 0.6$m, and (i) $H = 0.78$m as functions of the air pressure $P$. $V$ increases nearly linearly with a decrease of $P$ for various $H$'s and $D$'s; this increase is larger for larger $H$'s and smaller $D$'s, indicating a stronger air resistance for the two situations. The reader may note that $V$ increases from 2.94 m/s to 3.09 m/s for ethanol droplets with $D$ = 1.84mm released from $H$ = 0.5m (Fig. 8 (a)), but from 3.51 m/s to 3.76 m/s for droplets released from $H$ = 0.78m (Fig. 8 (c)); $V$ increases from 2.96 m/s to 3.10 m/s for water droplets with $D$ = 2.30mm released from $H$ = 0.5m (Fig. 8 (d)), but from 3.01 m/s to 3.10 m/s for droplets with $D$ = 2.74mm (Fig. 8 (g)).

Predictions using $C_d$ = 0 and Eqs (8) and (9) are unaffected by air pressure, indicating that they cannot account for pressure variations. Predictions based on Eqs (6) and (7) start from $P$ = 1 kPa in Fig.8 as the continuum model fails for rarified gases ($P \rightarrow 0$). The predictions produced by the first three models agree reasonably with the experiments, indicating their wide adaptability.

**5. Conclusion**

The velocity of the falling droplet just before its impact was determined experimentally for a range of droplet diameters, liquid density, release height and air pressure. The existing correlations for predicting droplet velocity were compared with the experimental results. It is shown that predictions produced by a dynamic model that utilizes the instantaneous values of the drag coefficient from two empirical correlations agree well with all the experiments in the range of parameters used in this study, providing a valuable tool for the determination of the droplet velocity used in impact studies.

**Acknowledgments**

This study was financially supported by the National Natural Science Foundation of China under Grant No. 12072032, the National Key R&D Program of China under Grant No. 2018YFF0300804, and 111 Project under Grant No. B16003.

*hjgizq@bit.edu.cn